# Self-organized annealing in laterally inhibited neural networks shows power law decay

Frank Emmert-Streib[1]
Institute for Theoretical Physics, University of Bremen
Otto-Hahn-Allee, 28334 Bremen
E-mail: fes@stowers-institute.org

*Abstract*—In this paper we present a method that assigns to each layer of a multilayer neural network, whose network dynamics is governed by a noisy winner-take-all mechanism, a neural temperature $\beta^{-1}$. This neural temperature is obtained by a least mean square fit of the probability distribution of the noisy winner-take-all mechanism to the distribution of a softmax mechanism, which has a well defined temperature as free parameter. We call this approximated temperature resulting from the optimization step the neural temperature. We apply this method to a multilayer neural network during learning the XOR-problem with a Hebb-like learning rule and show that after a transient the neural temperature decreases in each layer according to a power law $\beta^{-1} \sim t^{-\gamma}$. This indicates a self-organized annealing behavior induced by the learning rule itself instead of an external adjustment of a control parameter as in physically motivated optimization methods, e.g., simulated annealing.

## 1. Introduction

During the last decades research in the field of computational neuroscience has made considerable progress towards an understanding of the brain. The investigations extend from the behavior of a single neuron by Hodgkin and Huxley [8] to the interaction of all neurons in the brain whose activity can be visualized by brain imaging methods like fMRI [11]. Despite these achievements a general mathematical framework is still absent.

A major problem in dealing with a complex adaptive system like the brain is the vast number of free parameters, e.g., the synaptic weights between adjacent neurons, which has to be adapted during life in a meaningful way to ensure the survival of an animal. The problem with the adaptation of the synaptic weights is not only that there is a huge amount of synapses, but also that there is no central processing unit in the brain, which assigns these values. Instead the brain is organized according to local rules for the synaptic modifications as Hebb postulated already in 1949 [6]. Hence, the analogy between a serial von Neumann computer architecture omnipresent in our desktop computers, which is organized in a central way, and the brain breaks completely down in this point.

In this paper we address the question whether there is a variable in a neural network that reflects the performance of the network's output despite its local and decentral working mechanism. We demonstrate that one can assign to a laterally inhibited multilayer neural network an auxiliary variable in the form of an approximated temperature whose course reflects the error of the network during the learning of a problem. Moreover, its temporal course follows a power law decay.

---

[1] Present address: Stowers Institute for Medical Research, Bioinformatics, 1000 E. 50th Street, Kansas City, MO 64110, USA

This paper is organized as follows. In section 2 we analytically calculate the probability distribution of a noisy winner-take-all mechanism. This result is then used in section 3 to introduce a method that assigns an approximated temperature to a layer of laterally inhibited neurons. In the following section 4 we exemplarily demonstrate the application of this method during learning the XOR-problem in a three-layer neural network. The article concludes in section 5 with a summary of the results.

## 2. Analytical solution for the distribution of a noisy winner-take-all mechanism

A noisy winner-take-all mechanism, which is frequently used as network dynamics in neural network modeling [1,2,10], is a selection mechanism in a layer of a feed-forward network [7]. Mathematically it is defined as

$$h_i' = h_i + \eta_i$$
$$i_{max} = \arg\max_i(h_i') \quad (1)$$
$$x_i = \begin{cases} 1, & i = i_{max} \\ 0, & i \neq i_{max} \end{cases}$$

Here $h_i = \sum_j w_{ij} x_j$ is the inner field of neuron $i$ in the layer that is calculated via the input neurons $x_j$ of the preceding layer, and $\eta_i$ is noise drawn from a probability distribution $p_\eta$ independently for each neuron. The neuron with the highest inner field $h'_i$ after adding noise is selected to be active $x_{i_{max}} = 1$, all other neurons are inactive $x_{i \neq i_{max}} = 0$. For the following considerations it is useful to order the inner fields according their values in increasing order and renumber them new in a consecutive way from low to high values. One can now ask the question, what is the probability $p_1$ that neuron 1, the neuron with the lowest inner field after renumbering, is selected to be active via equations (1). The case $\eta_i = 0$ leads to the deterministic winner-take-all mechanism, which always selects the neuron with the highest inner field with probability 1 and all other neurons with probability 0. For $\eta_i \neq 0$ equations (1) define a stochastic process and the answer to this question is less obvious. In the rest of this section we analytically calculate the probability distribution for a special case to this general question, when $p_\eta$ is the equal distribution in $[0, \eta_{max}]$. We introduce an auxiliary variable $H_i = h_i + \eta_{max}$ which is the maximum value of the inner field after adding the highest possible noise value. We formalize the question disposed above and specify the probability that neuron i is selected by

$$p_i = P(h_i' > h_j', \forall j \neq i, p_h) \quad (2)$$

The condition for the selection of neuron $i$ is to have the highest inner field after adding noise from the noise distribution $p_\eta$ in comparison with all other inner fields. Expression (2) can be evaluated by weighting the probability density $\rho_i(H)$ for neuron $i$ with the probabilities $P(x_j \leq H)$ that the neurons $j \neq i$ are not chosen by integration from the value of the highest inner field $h_n$ to the maximal value which is reachable for neuron i.

$$p_i = \int_{h_n}^{H_i} r_i(H) P(x_1 \leq H) \mathbf{K} P(x_n \leq H) dH \quad (3)$$

Here $P(x_j \leq H)$ is the distribution function of neuron j

$$P(x_j \leq H) = \int_0^H r_j(H') dH'$$
$$r_j(H) = \frac{1}{h_{max}} \Theta(H - h_j) \Theta(H_j - H) \quad (4)$$

which gives for the equal distribution $\rho_j(H)$

$$P(x_j \leq H) = \begin{cases} 0 & : \quad H < h_j \\ \dfrac{H - h_j}{h_{max}} & : \quad h_j \leq H \leq H_j \\ 1 & : \quad H_j < H \end{cases} \quad (5)$$

This is the general formulation for n neurons. Due to the case decisions in (5) for every neuron $j \neq i$ in (4) the solution (3) for neuron i is a composition of these case decisions. We give the solutions for n=3 neurons explicitly because in the results section we apply these solutions to a three layer feed-forward network to learn to the XOR-Problem, which has three neurons in the second layer.

The probability that neuron 1 (the neuron with the lowest inner field) is chosen is given by

$$p_1 = \begin{cases} 0: & H_1 \leq h_3 \\ p_1^1: & H_1 > h_3 \end{cases} \qquad (6)$$

with

$$p_1^1 = \frac{1}{h_{max}^3}\left[\frac{H_1^3}{3} - \frac{H_1^2(h_2+h_3)}{2} + H_1 h_2 h_3 - \frac{h_3^3}{3} + \frac{h_3^2(h_2+h_3)}{2} - h_2 h_3^2\right] \qquad (7)$$

It is clear, that the maximal inner field $H_1$ has to be larger than $h_3$ otherwise there is no possibility to select this neuron. The probability for neuron 2 is given by

$$p_2 = \begin{cases} 0: & H_1 \leq h_3; H_2 \leq h_3 \\ p_2^1: & H_1 \leq h_3; H_2 > h_3 \\ p_2^2: & H_1 > h_3; H_2 > h_3 \end{cases} \qquad (8)$$

with

$$p_2^1 = \frac{1}{h_{max}^2}\left[\frac{H_2^2}{2} - H_2 h_3 + \frac{h_3^2}{2}\right]$$

$$p_2^2 = \frac{1}{h_{max}^3}\left[\frac{H_1^3}{3} - \frac{H_1^2(h_1+h_3)}{2} + H_1 h_1 h_3 - \frac{h_3^3}{3} + \frac{h_3^2(h_1+h_3)}{2} - h_1 h_3^2\right] \qquad (9)$$

$$p_2^3 = \frac{1}{h_{max}^2}\left[\frac{H_2^2}{2} - H_2 h_3 - \frac{H_1^2}{2} + H_1 h_3\right]$$

The result for $p_3$ is

$$p_3 = \begin{cases} 1: & H_1 \leq h_3; H_2 \leq h_3 \\ p_3^1 + p_3^2: & H_1 \leq h_3; H_2 > h_3 \\ p_3^3 + p_3^4 + p_3^5: & H_1 > h_3; H_2 > h_3 \end{cases} \qquad (10)$$

The auxiliary functions are given by

$$p_3^1 = \frac{1}{h_{max}^2}\left[\frac{H_2^2}{2} - H_2 h_2 - \frac{h_3^2}{2} + h_2 h_3\right]$$

$$p_3^2 = \frac{h_3 - h_2}{h_{max}}$$

$$p_3^3 = \frac{1}{h_{max}^3}\left[\frac{H_1^3}{3} - \frac{H_1^2(h_1+h_2)}{2} + H_1 h_1 h_2 - \frac{h_3^3}{3} + \frac{h_3^2(h_1+h_2)}{2} - h_1 h_2 h_3\right] \qquad (11)$$

$$p_3^4 = \frac{1}{h_{max}^2}\left[\frac{H_2^2}{2} - h_2 H_2 - \frac{H_1^2}{2} + h_2 H_1\right]$$

$$p_3^5 = \frac{h_3 - h_2}{h_{max}}$$

To check our analytical results (6),(8) and (10) we compared them with numerical simulations and obtained excellent correspondence [4].

In the next section we will use our results (6)-(11) to derive an approximated temperature for the noisy winner-take-all mechanism.

## 3. Approximated Temperature for the noisy winner-take-all mechanism

The motivation for the introduction of an approximated temperature for a noisy winner-take-all mechanism comes from two observations. First, the adaptation of the synaptic weights of a neural network is an optimization process with the goal to learn a given mapping. The optimization process is controlled by the learning rule used to adapt the synaptic weights. If one uses a softmax or noisy winner-take-all mechanism to introduce noise in the system, the values of the inner fields $h_i$ of the neurons organize during the learning process in a way that the influence of the noise is reduced. This results in a decreasing failure rate of the neural network during learning, until the system has eventually learned the mapping completely. However, the temperature like parameter $\beta^{-1}$ of the softmax mechanism or $\eta_{max}$ of the noisy winner-take-all mechanism was not changed during this process at all. Second, simulated annealing [9], a physically motivated optimization method, reduces gradually during the optimization process a temperature (or temperature like parameter when used in a non physical context) until the global (or for non trivial problems in reality a local) minimum is reached. These two observations are only compatible by recognizing that in the former case $\beta^{-1}$ and $\eta_{max}$ do not reflect the optimization state of the system itself as in simulated annealing but describe the perturbation of the system.

Now, the question arises, is there a parameter in a neural network that can be associated with a parameter as in simulated annealing, which time course reflects the progress in the optimization problem, and how is it obtained? As a possible answer to this question we propose to following approach. We use a noisy winner-take-all mechanism as network dynamics in the neural network and compare the corresponding probability distribution with the probability distribution of the softmax mechanism, which has a temperature like parameter. In principle, we use the softmax mechanism as virtual network dynamics.

More formally, we obtain the temperature like parameter $\beta^{-1}$ from a comparison of the softmax [12]

$$p_i^{soft} = \frac{\exp(bh_i)}{Z}$$
$$Z = \sum_i \exp(bh_i)$$
(12)

and the noisy winner-take-all mechanism (1), whose probability distribution is given by (6)-(11), by a minimization of the mean square error $E_{ms}$ between both distributions.

$$E_{ms} = \frac{1}{2}\sum_i \left(p_i^{nwta} - p_i^{soft}\right)^2$$
(13)

Numerical results show, that this is equivalent to the condition $p_n^{soft} = p_n^{nwta}$ for the component with the largest h value[2]. The numerical difference between both conditions is less then 1%. Hence, for numerical simplicity we determine $\beta^{-1}$ according to this condition, because we have to evaluate (13) at each time step during the learning process of the neural network. If we apply this method to a layer of a neural network we call the obtained approximated temperature the neural temperature of this layer. It is clear, that different layers in a multilayer neural network can have different neural temperatures, because each layer has its own distribution $p^{nwta}$ and hence, its own approximated temperature. We want to mention, that there are several other information theoretic measures available that allow to obtain $\beta^{-1}$ as a result from an optimization process, e.g. maximum entropy or relative entropy (Kullback Leibler distance). In an upcoming work we will compare some of these measures and discuss this point in detail.

In figure 1 we show an example for the inner fields $h_1=0.0$, $h_2=0.8$ and $h_3=1.0$ to demonstrate, that our optimization criteria works well. The distribution of the noisy winner-take-all mechanism, which was calculated analytically according to (6)-(11) in dependence of the noise $\eta$, is shown in full lines and the distribution of the softmax mechanism in dashed lines. One can see, that the difference between the pairwise components of the distributions, are moderate for all noise values.

---

[2] Remember, that we numbered the inner fields h in acceding order. Hence, $h_n$ is the largest inner field.

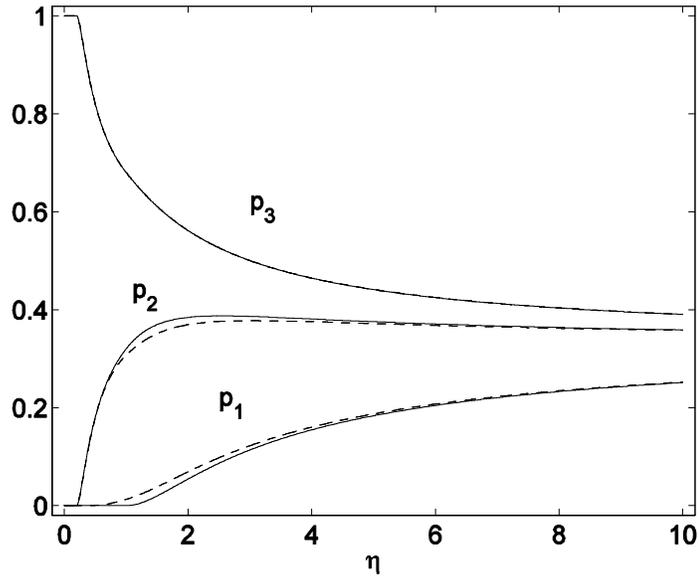

**Figure 1. Comparison of the distributions $p_i$ of the noisy winner-take-all (full lines) and the softmax mechanism (dotted lines) in dependence of noise $\eta$ for** $h_1=0.0$, $h_2=0.8$ and $h_3=1.0$**.**

## 4. Results

In this section we apply our method introduced in the last section to a multilayer feed-forward neural network [7] while learning a mapping to monitor the course of the neural temperature during the learning process. We use a network with three layers consisting of 3/3/2 neurons in the input/hidden/output layer and learn the XOR mapping. One input neuron serves as bias and is set permanently to 1 to prevent the case of zero activity in the network [10]. We use binary neurons and the network dynamics is given by a noisy winner-take-all mechanism. The connections between adjacent layers are all to all [10].

The learning rule we apply for the adjustment of the synaptic weights was recently introduced by the author in [4,5]. The main idea of this learning rule is to assign to each neuron in the network one additional degree of freedom in the form of a so called neuron counter $c_i$. The dynamics of these neuron counters is determined by the performance of the network itself. This is realized by a reinforcement signal r which is feed back to the neuron counters after the presentation of an input pattern assigning by r=1 a correct and r=-1 a wrong network output.

$$c_i \rightarrow c_i' = \begin{cases} \Theta, & c_i - r > \Theta \\ c_i - r, & \Theta \geq c_i - r \geq 0 \\ 0, & 0 > c_i - r \end{cases} \quad (14)$$

Here $\Theta$ is a positive integer, which has the meaning of the memory length of the neuron counters. Equation (14) is only applied to the neurons that were active for the last presented input pattern. The other neuron counters remain unchanged. Similar to [1,2,10] only active synapses involved in the signal processing of the last pattern can be updated if the output of the network was wrong. In this case the neuron counters are used to evaluate a stochastic update condition for these synapses. If the stochastic update condition

$$p_c < p_{d_{ij}}^r \quad (15)$$

is fulfilled the synapse is updated by

$$w_{ij} \rightarrow w_{ij}' = w_{ij} - d \quad (16)$$

The stochastic update condition (15) is obtained by calculating the approximated synapse counter $d_{ij}=c_i+c_j$ for all active synapses. We call $d_{ij}$ approximated synapse counter, because Klemm et.al. [10] introduced a learning rule for neural networks, which was based on synapse counters. That means, Klemm et al. assigned one additional degree of freedom to each synapse in form of a synapse counter, instead of a neuron counter in our learning rule.

Then the probability $p^r_{d_{ij}}$ is assigned from the rank ordering distribution

$$P^r_k \sim k^{-t}, t \in \Re^+ \\ k \in \{1, \mathbf{K}, 2\Theta + 3\} \qquad (17)$$

by the mapping k=2Θ+3 - $d_{ij}$ and the coin probability $p_c$ is drawn from the distribution

$$P_c \sim x^{-a}, a \in \Re^+ \\ x \in (0,1] \qquad (18)$$

A thorough discussion of this Hebb-like learning rule for neural networks and its biological interpretation can be found in [4,5]. We visualize in figure 2 the overall effect of the stochastic update condition (15) in dependency of the exponent α of the coin distribution on the update probability

$$P_{update} = P(p_c < p^r_{d_{ij}}) \qquad (19)$$

for the synapses by calculating explicitly the update probability for these cases. The higher the exponent α is the higher the update probability and vice versa. Hence, α is a parameter that controls the update frequency. For α→∞ the active synapses are always updated if r= -1. This would eliminate completely the effect of the neuron counter $c_i$ (14) which idea is to introduce a mean failure rate for each neuron on which the update decision is based.

Now we apply our method from the previous section to a three-layer neural network and calculate for the second (hidden) and third (output) layer the neural temperature during learning the XOR-problem. For the following simulations we used δ=1.0=const. for the synaptic modification, τ=2.0 for the rank ordering (17) and α =1.2 for the coin distribution (18). The noise level was chosen to be $\eta_{max}$=0.45. In all simulations the synaptic weights $w_{ij}$ are initially i.i.d. chosen from the interval [0,1] and the neuron counters $c_i$ are set to zero. The ensemble size for all simulations was N=5000.

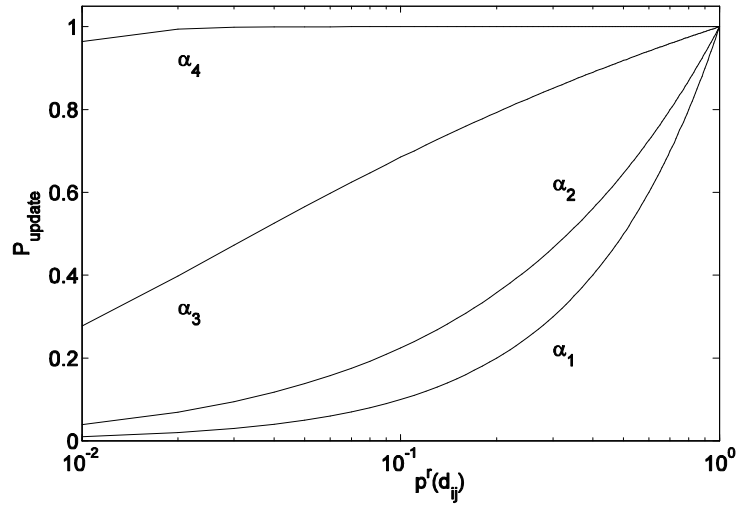

**Figure 2. Update probability** $P_{update} = P(p_c < p^r_{d_{ij}})$ **in dependency of** $p^r_{d_{ij}}$ **and the exponent of the coin distribution α for** $α_1$=0.0, $α_2$=0.4, $α_3$=1.2 **and** $α_4$=5.0.

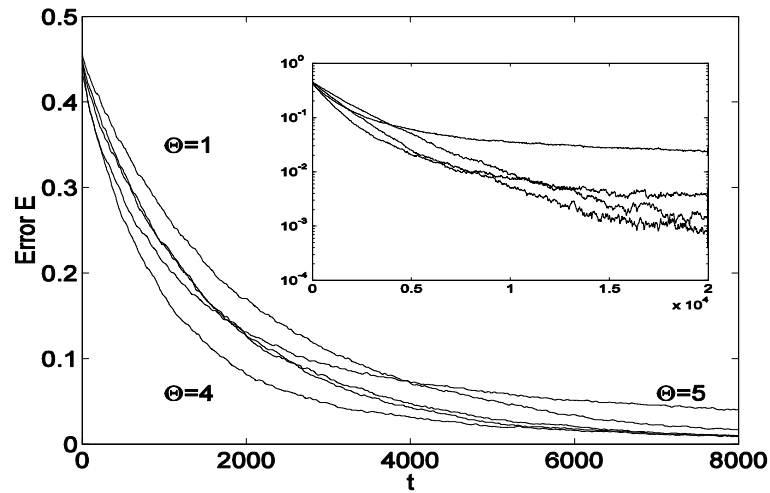

**Figure 3. Time course of the mean ensemble error E in dependence of the neuron counters Θ. The dependence of Θ in the inner figure is from top to bottom, Θ ={5,4,1,2}.**

First of all we show in figure 3 the mean ensemble error E(t) to demonstrate that our stochastic Hebb-like learning rule can learn the XOR-problem in the given network topology, because all curves are up to t ~ 8000 below an error of 5%. The definition of the mean ensemble error E(t) is given by

$$E(t) = \frac{1}{N}\sum_{i=1}^{N} e_i(t)$$
$$e_i(t) \in \{0,1\}$$
(20)

Here $e_i(t) \in \{0,1\}$ is the individual network error that indicates if the output of network $i \in \{1,\ldots,N\}$ at time step t was right $e_i(t)=0$ or wrong $e_i(t)=1$.

The dependency of the neuron counter values Θ for the given parameter configuration is moderate, but visible. The inner figure shows in a half-logarithmic plot the progress of learning up to $t=2*10^4$ time steps to demonstrate that the optimal parameter configuration depends on the time scale as can be seen by the intersections between different learning curves.

During the learning process of the XOR-problem we apply our method from section 3 at each time step to obtain the neural temperature $\beta^{-1}$ for the second (hidden) and third (output) layer. The upper figure 4 shows in a double-logarithmic plot the mean neural temperature for layer 2 obtained after ensemble averaging. One can see that after a transient, which is given in table 1, the long time annealing behavior of the mean neural temperature follows a power law $\beta^{-1} \sim t^{-\gamma}$ with different exponents for different neuron counter values. Additionally, the order from high to low of the mean ensemble errors E(t) in figure 3 corresponds to the order of the mean neural temperatures from high to low temperatures in the upper figure 4 after the transient. The time course of the neural temperature establishes a connection between the neural and behavioral level of description because the mean neural temperature indicates, without knowledge of the mean ensemble error, if learning in the neural network takes place. E.g., this can be seen in analogy to simulated annealing, however with the difference, that the neural temperature emerges self-organized by the learning rule of the network itself and not by manually tuning an external control parameter.

In the lower figure 4 the corresponding results for the mean neural temperature in layer 3 during the learning process are shown. These results confirm our observations in layer 2. The only differences between the two layers are the quantitative values of the exponents γ given in table 1. Simulations for other parameter values of τ, α and $\eta_{max}$ confirm our result, namely that the neural temperature anneals during learning according to a power law. Moreover, this holds also for learning rules other then ours, e.g. for the learning rule proposed by Klemm et.al. [10] indicating the general character of this result. A comparison of these results will be given in a future work.

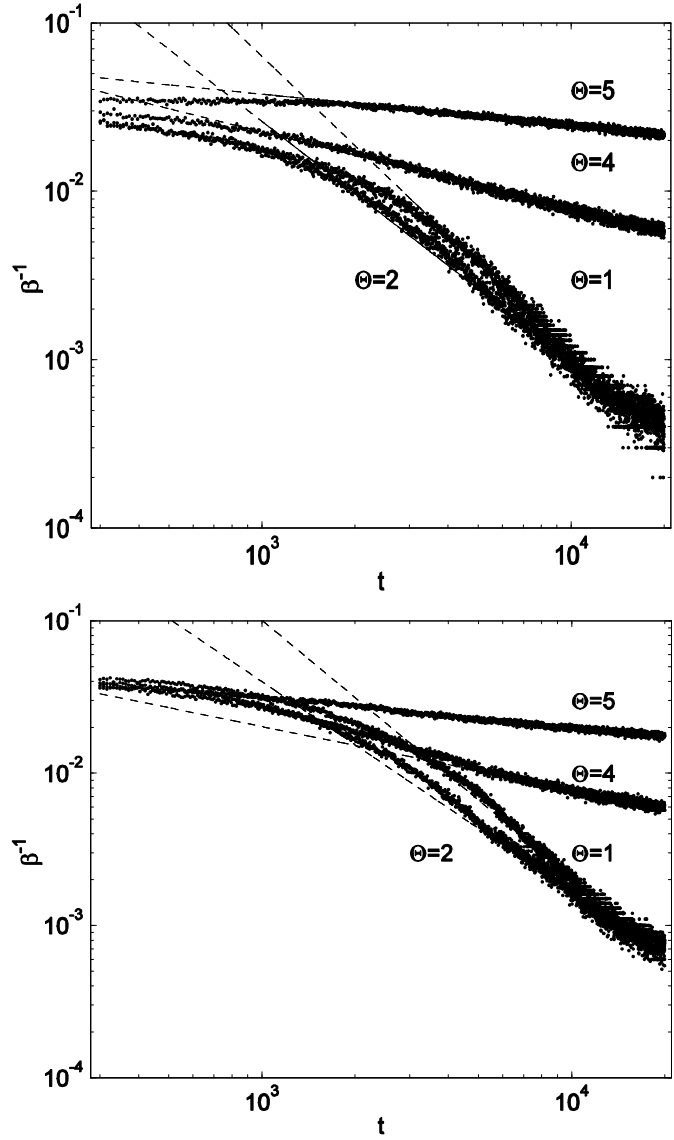

**Figure 4.** Time course of $\beta^{-1}$ in dependence of $\Theta$. The upper figure corresponds to layer 2 the lower one to layer 3. The straight lines are fitted to $\beta^{-1}$ from time points which are given in table 1.

**Table 1.** Exponents of the power law $\beta^{-1} \sim t^{-\gamma}$ for layer 2 and 3 during learning the XOR-problem. $t_{Fit}$ gives the time point from which on the power law was fitted.

|  | $g_2$ | $t_{Fit}$ | $g_3$ | $t_{Fit}$ |
|---|---|---|---|---|
| $\Theta = 1$ | 1.76±0.003 | 5000 | 1.69±0.002 | 6000 |
| $\Theta = 2$ | 1.42±0.003 | 5000 | 1.38±0.003 | 6000 |
| $\Theta = 3$ | 1.44±0.002 | 2000 | 1.26±0.002 | 5000 |
| $\Theta = 4$ | 0.46±0.001 | 2000 | 0.41±0.001 | 5000 |
| $\Theta = 5$ | 0.18±0.002 | 2000 | 0.18±0.001 | 2000 |

## 5. Conclusions

We introduced in this article a new general method, which assigns to a noisy winner-take-all mechanism, often used in neural networks as lateral inhibition, a neural temperature and, hence, connects the effective noise perturbing the system explicitly to a temperature. This method was used to investigate the behavior of the neural temperature during a learning process of a three-layer neural network. The neural network was trained to learn the XOR-problem according to a Hebb-like learning rule.

Our simulations reveal that the course of the neural temperature corresponds to the network's error, which is objectively measurable by the mean ensemble error. Moreover, it obeys power law decay in both layers indicating that network activity reaches a critical state during the learning process. However, the exponent of the power law is not universal for the learning rule but depends on its constituting parameters. We want to emphasize that the annealing behavior of the neural temperature occurs self-organized by the learning rule of the network itself and not by manually tuning an external control parameter. This is in contrast to physically motivated optimization methods, e.g., simulated annealing [9], where it is necessary to have a problem specific temperature scheduling.

These observations raise some speculations concerning the working mechanism of the real brain. Does the brain reach during learning a critical state or is it during the learning process in one? This question was in similar form already addressed by P. Bak and D. Chialvo [1,3]. We neither know the answer to this question nor are we sure that our model[3] is sufficiently complex to address this question. However, we have the strong feeling that criticality and self-organization play a dominating role in the organization of the brain, as P. Bak already pointed out and see in our results a contribution on the way to demonstrate this.

**Acknowledgment:** We would like to thank Rolf D. Henkel, Jens Otterpohl, Klaus Pawelzik, Roland Rothenstein and Helmut Schwegler for fruitful discussions and Tom Bielefeld and Earl Glynn for carefully reading the manuscript.

---

[3] Our model is the same as used by [1], which was called by them the mini brain model.

**Frank Emmert-Streib** obtained his *Diploma* in Theoretical Physics in 1998 form the University of Siegen (Germany) and his Ph.D. in Theoretical Physics from the University of Bremen (Germany) in 2003. He is currently a postdoctoral research associate in Bioinformatics at the Stowers Institute for Medical Research (USA). His research interests include Computational Biology, Machine Learning and Systems Biology. (Home page: http://www.bio-complexity.com)